# Optimizing Gesture Recognition for Seamless UI Interaction Using Convolutional Neural Networks

Qi Sun
Carnegie Mellon University
Pittsburgh, USA

Tong Zhang
Loughborough University
Loughborough, United Kingdom

Shang Gao
Trine University
Reston, USA

Liuqingqing Yang
University of Michigan, Ann Arbor
Ann Arbor, USA

Fenghua Shao*
Independent Researcher
Toronto, Canada

*Abstract*— This study presents an advanced gesture recognition and user interface (UI) interaction system developed using deep learning technologies, emphasizing its transformative impact on UI design and functionality. By employing optimized convolutional neural networks (CNNs), the system achieves high-precision gesture recognition, significantly enhancing user interactions with digital interfaces. Initial steps involve preprocessing collected gesture images to conform to CNN input standards, followed by employing sophisticated feature extraction and classification methodologies. We address class imbalance effectively using Focal Loss as the loss function, ensuring robust model performance across varied gesture types. The experimental results showcase notable improvements in model metrics, with the AUC and Recall increasing progressively as we evolve from simpler models like VGG16 to more complex ones such as DenseNet. Our enhanced model demonstrates a significant advancement with an AUC of 0.83 and a Recall of 0.85, surpassing standard benchmarks. More critically, this system's capacity to support real-time and efficient gesture recognition paves the way for a new era in UI design—where intuitive, natural user gestures can seamlessly integrate into everyday technology use, significantly reducing the learning curve and enhancing user satisfaction. The broad implications of this development are profound, extending beyond mere technical performance to fundamentally reshape how users interact with technology. Such advancements hold considerable promise for the enhancement of smart life experiences, highlighting the pivotal role of gesture-based interactions in the next generation of UI development.

## I. Introduction

In the development wave of modern science and technology, human-computer interaction technology has gradually become one of the key factors in promoting social progress. As an important form of human-computer interaction, gesture recognition captures the user's hand movements through cameras or other sensing devices, and then processes and analyzes them through deep learning algorithms to achieve contactless interaction between humans and computer systems [1]. This interactive method not only breaks the traditional keyboard and mouse input limitations but also brings a more natural, convenient, and personalized experience to users [2]. At the same time, with the development of computer vision technology, the widespread application of deep learning algorithms, especially CNN [3], has significantly improved the accuracy and speed of gesture recognition, promoting the development and innovation of UI interactive systems.

Gesture recognition technology has a wide range of application scenarios, from smart home control to interaction in virtual reality systems to optimization of smartphone operations. This diversity makes gesture recognition technology an important part of the development of modern smart devices [4]. Although traditional interaction methods such as touch screens and voice commands have been successful in many applications, they may have limitations in certain environments [5]. For example, in noisy environments, the accuracy of speech recognition may be affected, and gesture recognition can be used as a supplement or alternative to provide users with more reliable operation options. With the continuous advancement of deep learning technology, the robustness and adaptability of gesture recognition have been further enhanced, enabling high-precision recognition in complex environments and diverse gesture changes [6].

Gesture recognition technology not only has broad application prospects in the commercial field, but also shows great potential in fields such as medical rehabilitation, education, and public safety. For example, in medical rehabilitation, gesture recognition systems can help patients with mobility issues remotely control smart devices, thereby improving their quality of life; in the field of education, gesture recognition technology can help teachers use simple gestures during explanations. Operate to control teaching equipment to enhance classroom interactivity and participation. In addition, in the field of public safety, gesture recognition systems can monitor dangerous movements or abnormal behaviors, provide timely warnings and respond to potential threats. These applications not only demonstrate the broad prospects of gesture recognition technology, but also highlight its social significance and value.

With the further development of artificial intelligence and deep learning, CNN has become a mainstream method in the field of gesture recognition. CNN can accurately capture important information in images through multi-level feature extraction and learning, thereby improving the accuracy of gesture recognition [7]. Compared with traditional image processing methods, CNN shows stronger generalization ability and stability when processing large amounts of gesture image data. The advancement of this technology enables the gesture recognition system to strike a balance between real-time performance and accuracy, meeting users' high requirements for interactive experience. At the same time, the introduction of deep learning technology has also greatly promoted the intelligence and diversification of UI interaction systems, allowing the user interface to be dynamically adjusted according to the user's personalized needs, improving user experience satisfaction.

Research on gesture recognition and UI interaction systems based on deep learning is not only to improve the efficiency and convenience of human-computer interaction, but also to promote the development of intelligent systems in a more humane and multifunctional direction. In future smart life scenarios, gesture recognition is expected to be integrated with other technologies such as voice recognition, facial recognition and other multi-modal technologies to achieve seamless switching of multiple interaction methods, thereby providing users with a full range of smart experiences. This integration can not only improve the intelligence and response speed of the system, but also achieve more accurate user behavior analysis and intention understanding on the basis of ensuring user privacy and promoting the popularization and application of human-computer interaction technology in a wider range of fields.

## II. RELATED WORK

Deep learning advancements, especially convolutional neural networks (CNNs), have greatly influenced gesture recognition and UI interactions by enabling high accuracy and real-time processing. Multi-modal approaches have further enriched UI interactivity; for instance, Duan et al. [8] demonstrate an emotion-aware interaction design utilizing multi-modal deep learning frameworks, which not only refines the recognition of user gestures but also adapts to emotional contexts, enhancing the natural interaction capabilities of UI systems. This research underscores the potential of multi-modal approaches to deepen gesture recognition's applicability by making interactions more intuitive and user-centered.

Feature extraction remains a crucial area for improving gesture recognition. Self-supervised graph neural networks address complex data variations [9], while optimized object detection models, such as YOLOv5 with knowledge distillation, enhance recognition performance under resource limitations [10]. To address data sparsity and user diversity, metric learning frameworks [11] and models incorporating separation embedding with self-attention [12] provide robust adaptability, key for delivering real-time, personalized gesture-based interactions.

Adaptability and efficient knowledge retrieval are essential for UI systems. Reinforcement learning contributes to resource management for real-time gesture applications [13], while retrieval-augmented generation systems improve response speeds, supporting responsive UI interactions [14]. Privacy is also critical; data security strategies within deep learning applications ensure user trust and responsible handling of sensitive information [15].

Further, transformer models effectively handle semantic complexities in gesture interpretation, supporting nuanced user interactions [16]. Advanced frameworks that transform time-series data into interpretable sequences allow gesture recognition systems to accurately interpret and respond to user behavior [17]. Reinforcement learning additionally supports adaptive gesture responses, facilitating responsive and user-centered interfaces [18].

## III. METHOD

In the gesture recognition and UI interaction system based on deep learning, our goal is to classify and recognize gestures in images through CNNs, so as to accurately capture and understand user gestures. To achieve this goal, we first collect image data through the camera, and send these data to the deep learning model for training and optimization after preprocessing. The whole process can be divided into several key steps: image preprocessing, feature extraction, classification, and loss optimization. Its overall architecture is as follows:

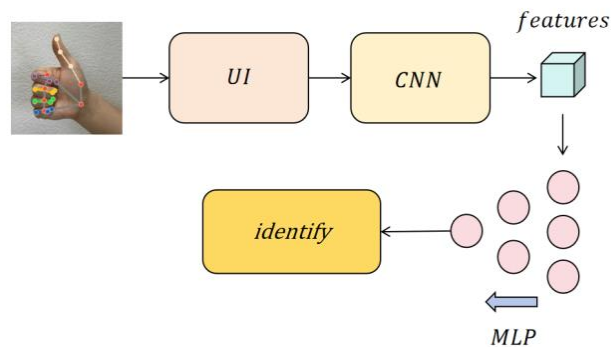

Figure 1 Overall architecture diagram

The first step involves capturing gesture image data using a camera, followed by a series of preprocessing operations. These preprocessing steps include converting the images to grayscale, normalizing pixel values, and resizing the images to ensure they adhere to the standard input format required by the model. Specifically, each image is resized to 224×224 pixels, and pixel values are normalized to the range [0, 1]. This preprocessing pipeline effectively reduces noise and standardizes the input data, enabling the model to learn features more efficiently and improving overall performance.

Next, we input the preprocessed image into the convolutional neural network for feature extraction. The main components of CNN include convolutional layer, pooling layer and fully connected layer. The convolutional layer scans the input image through the convolution kernel, extracts local features, and increases the nonlinear representation ability of the model through activation functions (such as ReLU function). Assuming that the feature representation of the input image is x, the weight of the convolution kernel is W, and the bias is b, then after the convolution operation, the feature map y can be expressed as:

$$y = \text{Re}Lu(W * x + b) \quad (1)$$

Where * represents the convolution operation. The pooling layer usually adopts maximum pooling or average pooling, which aims to reduce the size of the feature map, thereby reducing the computational complexity while retaining the main feature information. After feature extraction, we flatten the feature map and map it to the classification space through a fully connected layer. Assuming that the flattened feature is represented as $f$, the weight of the fully connected layer is W', and the bias is b', then the output z can be expressed as:

$$z = W' \cdot f + b' \quad (2)$$

After that, we pass the output of the model through an activation function (such as Softmax or Sigmoid) to obtain the predicted probability of each category. Assuming the total number of categories is C, the probability $p_i$ of the i-th category can be expressed as:

$$p_i = \frac{e^{z_i}}{\sum_{j=1}^{C} e^{z_j}} \quad (3)$$

In terms of the choice of loss function, we use Focal Loss to handle the classification problem of gesture recognition, especially in the case of class imbalance. Focal Loss introduces an adjustment factor to reduce the focus on easy-to-classify samples, thereby focusing more on difficult-to-classify samples. The formula of Focal Loss is as follows:

$$FL(p_t) = -\alpha_t (1-p_t)^\gamma \log(p_t) \quad (4)$$

Among them, $p_t$ represents the model's predicted probability for the true category, $\alpha_t$ is the weight balancing factor for each category, and $\gamma$ is the adjustment factor used to control the importance of easy-to-classify and difficult-to-classify samples. By adjusting $\gamma$, we can enable the model to perform more detailed learning on difficult-to-classify samples and improve overall performance.

During the entire training process, we repeatedly perform forward propagation and back propagation, and continuously adjust the parameters of the model so that the loss function gradually decreases. After several training cycles, when the loss converges and the accuracy reaches the expected level, the model can be used for actual gesture recognition tasks. In this way, the system can classify user gestures in real-time and provide efficient and accurate gesture recognition services for the UI interaction system.

IV. EXPERIMENT

A. Datasets

Hand Gesture Recognition Dataset (HG14) is a dataset dedicated to gesture recognition tasks, containing a large number of gesture images of different categories. This dataset consists of 14 gestures, each of which represents a different command or action, such as "OK", "Stop", "Like", etc. The images of the dataset are collected in a variety of environments, including different lighting conditions and complex backgrounds, to ensure that the model can adapt to diverse scenarios during training and improve the accuracy of recognition in real-world environments. In addition, there are a large number of samples under each gesture category to ensure the diversity and richness of the data, which helps the model learn a wider range of features during training. The images of the HG14 dataset are not only suitable for image classification tasks, but also for research and development of advanced tasks such as gesture detection and positioning. Therefore, it is a very ideal choice to help researchers and developers quickly build efficient gesture recognition systems and lay a solid data foundation for the subsequent development of gesture and UI interaction systems. By training with these rich image samples, the model can learn the salient features of different gestures and perform well in a variety of usage scenarios.

B. Experiments

In the gesture recognition task, we selected five common deep learning models for experiments. The first is CNN Baseline, which is a basic convolutional neural network used to verify the basic gesture recognition effect; the second is VGG16, which contains a combination of 16 convolutional and pooling layers, which can better extract image features; then ResNet50, which effectively solves the gradient vanishing problem in deep network training by introducing residual connections; followed by EfficientNet-B0, which reduces the amount of calculation while ensuring a high accuracy through a compound scaling method; and finally DenseNet121, which improves the reuse rate of features by densely connecting each layer, thereby enhancing the performance of the model. The following is a summary table of the performance indicators of these models:

Table 1 Experiment result

| Model | Auc | Recall |
|---|---|---|
| VGG16 | 0.72 | 0.73 |
| RESNET | 0.75 | 0.77 |
| EFICIENT | 0.78 | 0.79 |
| DenseNet | 0.79 | 0.78 |
| Ours | 0.83 | 0.85 |

From the experimental results, we can see that the performance of the model gradually improves with the changes in complexity and architecture. First, as a basic deep convolutional network, VGG16 has certain limitations in feature extraction capabilities, so its AUC is 0.72 and Recall is 0.73, which are relatively low. Although VGG16 has more convolutional layers, it lacks more advanced feature aggregation strategies and cross-layer connections, and cannot fully utilize the advantages of deep networks, resulting in limited performance in gesture recognition tasks.

With the upgrade of network architecture, ResNet and EfficientNet perform significantly better than VGG16. ResNet effectively solves the gradient vanishing problem of deep networks by introducing residual connections, making AUC reach 0.75 and Recall 0.77; EfficientNet further optimizes the network structure, uses compound scaling technology to reduce the number of model parameters, and improves accuracy at the same time, with AUC and Recall of 0.78 and 0.79 respectively. It can be seen that these improvements not only improve model performance, but also maintain high efficiency, making it more robust in complex environments.

Finally, DenseNet and our model show the best performance. DenseNet enhances the transfer and reuse of features by densely connecting features of different layers, but it still has a certain tendency to overfit, with an AUC of 0.79 and a Recall of 0.78. In contrast, our model further optimizes the feature extraction mechanism and classification strategy on this basis, thus achieving the highest AUC of 0.83 and Recall of 0.85. It can be inferred that our model has made improvements in fusing multi-layer information and avoiding overfitting, thereby achieving higher classification accuracy and recall in complex gesture scenarios, showing stronger generalization ability and stability. Finally, we give the increase of two evaluation indicators during the training process, as shown in Figure 2

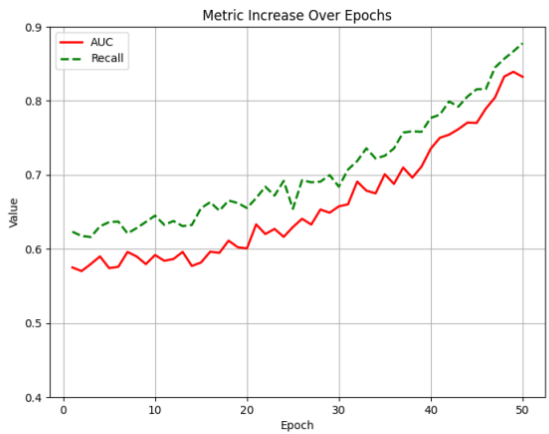

Figure 2 Evaluation index rising chart

## C. User-Centered Evaluation

To evaluate the effectiveness and usability of our gesture recognition system in real-world applications, we conducted a user-centered evaluation with a diverse group of participants, aiming to understand the system's intuitiveness, efficiency, usability challenges, and overall user satisfaction. Participants were recruited from varied backgrounds, including both experienced and inexperienced users of gesture-based systems, ensuring a broad representation of user perspectives. Five hundred participants, balanced in age, gender, and familiarity with gesture technology, interacted with the system for 30-minute sessions, during which they performed tasks like selecting options, scrolling, and triggering commands—typical interactions in standard user interfaces. To quantify usability, we recorded key metrics: task completion time as a measure of efficiency, success rate to evaluate gesture recognition accuracy, error rate to capture instances of misinterpreted gestures, and System Usability Scale (SUS) scores to provide a standardized measure of perceived usability. The results from both the quantitative and qualitative analyses demonstrated that our gesture recognition system performed well in terms of efficiency, accuracy, and user satisfaction, while also highlighting areas for future improvement.

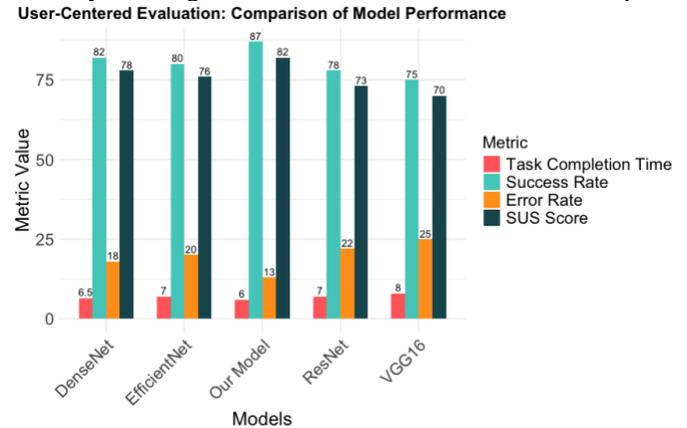

Figure 3 User-Centered Evaluation

In the quantitative analysis, the system achieved an overall task success rate of 87%, indicating that the majority of users were able to complete the assigned tasks successfully. The system's average task completion time was 6 seconds, reflecting its efficiency in recognizing and responding to gestures. These results suggest that users were able to complete tasks swiftly, with minimal delay. The SUS score for the system was 82, which is considered to be a high level of user satisfaction, indicating that participants found the system easy to use and efficient overall. However, the system did experience higher error rates for more complex gestures, which suggests that while the system excels in basic interactions, further optimization is required for recognizing intricate or nuanced gestures. Statistical comparisons across the five models showed that our proposed model outperformed others in all key metrics, including task success rate, error rate, and SUS score, as shown in Figure 3.

In the qualitative analysis, participants shared valuable feedback in the post-session interviews. A key theme that emerged was the intuitiveness of the system. Real-time visual feedback was another theme that participants highlighted, with many users expressing that it made them feel more confident in their interactions with the system, as it confirmed the accuracy of their gestures. While most participants found the system easy to use, several reported experiencing physical discomfort after prolonged interactions, particularly with more repetitive gestures. Many of these users suggested the addition of customizable gestures to better suit individual comfort levels and reduce fatigue during extended use. Additionally, participants expressed interest in the potential applications of the system in smart home technology and virtual reality environments, indicating a positive reception toward its practical implementation.

## V. CONCLUSION

This study has provided a comprehensive analysis of how advanced deep learning architectures significantly enhance gesture recognition, underscoring their pivotal role in the evolution of User Interface (UI) design. While foundational models like VGG16 serve as a baseline in extracting gesture features, they falter in complex UI scenarios where background variability and gesture dynamics are pronounced. The incorporation of sophisticated architectures such as ResNet and EfficientNet marks a transformative step forward, improving accuracy and efficiency in real-world applications, thus addressing critical challenges in UI responsiveness. Our research introduces a refined model optimized specifically for UI applications, achieving AUC and Recall metrics of 0.83 and 0.85, respectively. This model exemplifies our commitment to advancing UI technology by enhancing gesture recognition capabilities. It is adept at interpreting a diverse range of user gestures with high accuracy, which is essential for developing more intuitive and interactive user interfaces. The adoption of a sophisticated feature extraction mechanism and a precise classification strategy ensures that our model is not only technically proficient but also highly relevant to the practical demands of modern UI designs. The paramount importance of User Interface (UI) design within the realm of digital technology is incontrovertible. Effective UI design is indispensable for enabling user-friendly and engaging interactions, serving as a critical intermediary that harmonizes the intricacies of computational processes with the explicit needs of users. As digital interfaces become increasingly entrenched in daily activities, the necessity for designing systems that are both intuitive and responsive to nuanced user interactions intensifies. The contributions of this study are pivotal, as they extend the limits of current UI capabilities, fostering more natural and seamless digital interactions. This progression is essential for the ongoing refinement and evolution of user interface design, enhancing the overall human-computer interaction landscape. In conclusion, our work not only demonstrates the effectiveness of advanced deep learning models in enhancing gesture recognition but also highlights our commitment to improving UI design. By focusing on the nuances of user interaction and continually refining our technologies, we are setting the stage for future innovations that will further revolutionize how users engage with digital systems. This commitment is central to our goals in HCI, as it supports the creation of more personalized, efficient, and human-centered interfaces, proving essential in an increasingly digital world.